\documentclass[useASM, referee]{mn2e}
\usepackage{subfigure}
\input{psfig.sty}
\usepackage{graphicx}
\usepackage{lscape}
\begin{document}
\title[Power law relationship between GRB pulse width ratio and energy]{A test of the power law relationship between gamma-ray burst pulse width ratio and energy expected in
fireballs or uniform jets\footnote{Send offprint requests to:
Y.-P. Qin (E-mail: ypqin@ynao.ac.cn)}}
\date{2005 January 8}
\pubyear{????} \volume{????} \pagerange{2} \onecolumn
\author[Peng et al.]{Z.-Y. Peng$^{1,3}$, Y.-P. Qin$^{1,2}$, B.-B.
Zhang$^{1,3}$,
R.-J. Lu$^{1,2,3}$, L.-W. Jia$^{1,3}$, Z. -B. Zhang$^{1,3}$\\
$^1$National Astronomical Observatories/Yunnan Observatory,
Chinese Academy of Sciences,\\P. O. Box 110, Kunming 650011, China\\
$^2$Physics Department, Guangxi University, Nanning, Guangxi 530004, P. R. China\\
$^3$The Graduate School of the Chinese Academy of Sciences\\
 }
\pagestyle{plain}
\date{Accepted ????.
      Received ????;
      in original form 2005 April 18}
\pagerange{\pageref{firstpage}--\pageref{lastpage}} \pubyear{2005}
\maketitle \label{firstpage}
\begin{abstract}

A power law relationship between the pulse width and energy of
gamma-ray bursts (GRBs) was found by many authors. Recently, under
the assumption that the Doppler effect of the relativistically
expanding fireball surface (or in some papers, the curvature effect)
is important, Qin et al. showed that in most cases the mentioned
power law relationship would exist in a certain energy range and
within a similar range a power law relationship of an opposite trend
between the ratio of the rising width to the decaying width and
energy would be expectable for the same burst. We check this
prediction with two GRB samples which contain well identified
pulses. A power law anti-correlation between the full pulse width
and energy and a power law correlation between the pulse width ratio
and energy are seen in the light curves of the majority (around
65\%) of bursts of the two samples within the energy range of BATSE,
suggesting that these bursts are likely to arise from the emission
associated with the shocks occurred on a relativistically expanding
fireball surface. For the rest of the bursts, the relationships
between these quantities were not predicted previously. We propose
to consider other spectral evolutionary patterns or other radiation
mechanisms such as a varying synchrotron or Comptonized spectrum to
check if the observed relationships for these rest bursts can also
be accounted for by the Doppler model. In addition, we find that the
upper limits of the width ratio for the two samples do not exceed
0.9, in agrement with what predicted previously by the Doppler
model. The plateau/power law/plateau and the peaked features
predicted and detected previously by Qin et al. are generally
observed, with the exceptions being noticed only in a few cases.
According to the distinct values of two power law indices
$\alpha_{FWHM}$ and $\alpha_{ratio}$, we divide the bursts into
three subsets which are located in different areas of the
$\alpha_{FWHM}-\alpha_{ratio}$ plane. We suspect that different
locations of ($\alpha_{FWHM}$, $\alpha_{ratio}$) might correspond to
different mechanisms.

\end{abstract}


\section{Introduction}

Information of the dependence of the temporal profiles of
gamma-ray bursts (GRBs) on energy is important since it might
reflect the process of emission and suggest the form of radiation.
However, the variance of the GRB temporal profiles is very
enormous, and one cannot find two bursts having exactly the same
temporal and spectral developments (Nemiroff et al. 1993). Some
well-separated pulses which appear to have a fast rise and an
exponential decay (FRED) phases (see Fishman et al. 1994) are
generally believed to represent the fundamental constituent of GRB
light curves. Many investigations on modeling pulse profiles have
previously been made, and several flexible functions describing
the profiles of individual pulses based on empirical relations
were proposed (see e.g., Norris et al. 1996; Lee et al. 2000a,
2000b; Ryde et al. 2000, 2002). When fitted with these functions
many statistical properties of GRB pulses have been revealed.

In early statistical analysis, light curves of GRB pulses were
found to become narrower at higher energies (Fishman et al. 1992;
Link, Epstein, \& Priedhorsky 1993). Fenimore et al. (1995) used
the average autocorrelation function to study the average pulse
width, and showed that the average pulse width of many bursts is
well fitted by a power law of energy and the power law index is
about $-0.4$. Norris et al. (1996) also found that the average
pulse shape dependence on energy is approximately a power law,
consistent with the analysis of Fenimore et al. (1995). This was
confirmed by later studies (Norris et al. 1996, 2000; Costa 1998;
Piro et al. 1998; Nemiroff 2000; Feroci et al. 2001; Crew et al.
2003).

Many authors have attempted to interpret the light curves of GRBs in
the past few years (see, e.g., Fenimore et al. 1996; Norris et al.
1996; Norris et al. 2000; Ryde \& Petrosian 2002; Kocevski et al.
2003). It was suggested that the power law relationship could be
attributed to synchrotron radiation (see Fenimore et al. 1995; Cohen
et al. 1997; Piran 1999). Kazanas, Titarchuk, \& Hua (1998) proposed
that the relationship could be accounted for by synchrotron cooling
(see also Chiang 1998; Dermer 1998; and Wang et al. 2000). It was
suspected that the power law relationship might result from a
relative projected speed or a relative beaming angle (Nemiroff
2000). Phenomena such as the hardness-intensity correlation and the
FRED form of pulses were recently interpreted as signatures of the
relativistic curvature effect (Fenimore et al. 1996; Ryde \&
Petrosian 2002; Kocevski et al. 2003; Qin et al. 2004; Qin \& Lu
2005; Qin et al. 2005, hereafter Paper I). It is likely that the
observed difference between different channel light curves might
mainly be due to the energy channels themselves, owing to the
feature of self-similarity across energy bands observed (see, e.g.,
Norris et al. 1996). In other words, light curves of different
energy channels might arise from the same mechanism (e.g.,
parameters of the rest frame spectrum and parameters of the
expanding fireballs are the same for different energy ranges),
differing only in the energy ranges involved. This is what the
Doppler model (or in some papers, the curvature effect) predicts
(see Qin et al. 2004). The Doppler model is the model describing the
kinetic effect of the expanding fireball surface on the radiation
observed, where the variance of the Doppler factor and the time
delay due to different emission areas on the fireball surface (or
the spherical surface of uniform jets) are the key factors to be
concerned (for a detailed description, see Qin 2002 and Qin et al.
2004).

The observed gamma-ray pulses are believed to be produced in a
relativistically expanding and collimated fireball because of the
large energies and the short timescales involved. As shown in
Kocevski et al. (2003), when taking into account the curvature
effect, a FRED pulse can be expected. With their equations,
individual pulse shapes of a GRB sample were well characterized.

The formula of the Doppler model derived in details in Qin (2002) is
applicable to cases of relativistic, sub-relativistic, and
non-relativistic motions as no terms are omitted in the
corresponding derivation. With this formula, Qin (2003) studied how
emission and absorption lines are affect by the effect. Qin et al.
(2004) rewrote this formula in terms of the integral of the local
emission time, which is in some extent similar to that presented in
Ryde \& Petrosian (2002), where relation between the observed light
curve and the local emission intensity is clearly illustrated. Based
on this model, many characteristics of profiles of observed
gamma-ray burst pulses could be explained. Profiles of FRED pulse
light curves are mainly caused by the fireball radiating surface,
where emissions are affected by different Doppler factors and
boostings due to different angles to the line of sight, and they
depend also on the width and structure of local pulses as well as
rest frame radiation mechanisms. This allows us to explore how other
factors such as the width of local pulses affect the profile of the
light curve observed. Recently, Qin et al. (Paper I) studied in
details how the pulse width of gamma-ray bursts is related with
energy under the assumption that the sources concerned are in the
stage of fireballs. As revealed in Paper I, owing to the Doppler
effect of fireballs, it is common that there exists a power law
relationship between the full width at half-maximum ($FWHM$) and
energy and between $r_{FWHM}/d_{FWHM}$ and energy within a limited
range of frequency, where $r_{FWHM}$ and $d_{FWHM}$ are the $FWHM$
widths in the rising and decaying phases of the light curve,
respectively. They showed that, while emission of pulses over a
relativistically expanding fireball surface would lead to A power
law anti-correlation between the full pulse width and energy, it
would lead to a power law correlation between the ratio of the
rising width to the decaying width and energy. The power law range
and the corresponding index not only depend on the rest frame
radiation form but also on the observed peak energy (the range could
span over more than one to five orders of magnitudes of energy for
different rest frame spectra). The upper and lower limits of the
power law range can be determined by the observed peak energy
$E_{p}$. In cases when the development of the rest frame spectrum
could be ignored, a plateau/power law/plateau feature would be
formed, while in cases when the rest frame spectrum is obviously
softening with time, a peaked feature would be expected. In
addition, they found that local pulse forms affect only the
magnitude of the width and the ratio of widthes.

Although A power law anti-correlation between the pulse width and
energy was observed by many authors, it is unclear if a power law
correlation between the width ratio and energy could be detected in
the same source. First of all, we would like to check if the
expected power law relationships between $FWHM$ and energy and
$r_{FWHM}/d_{FWHM}$ and energy indeed hold for GRBs. When they hold,
how the two power-law indices are related? The primary goal of this
paper is to employ GRB samples to check in details these expected
relationships and to explore the possible relationship between the
two indices. In section 2, we present our sample description and
pulse fitting. The result are given in section 3. Discussion and
conclusions are presented in the last section.

\section{ Sample description and light curve fitting}

The first GRB sample we select comes from Kocevski et al. (2003),
where the bursts are found to contain individual FRED pulses. The
data are provided by the BATSE instruments on board the CGRO
spacecraft. The bursts of the sample they selected are from the
entire BATSE catalog with the criteria that the peak flux is greater
than 1.0 photons $cm^{-2}s^{-1}$ on a 256 ms timescale. They limited
the bursts to events with durations longer than 2 s. The sample
consists of 67 bursts. (For more details of the sample selection,
see Kocevski et al. 2003.) The second sample is presented in Norris
et al. (1999) which contains 66 single pulse GRBs. They performed a
several-step program, starting with the largest available sample of
bursts and decimating the sample according to criteria designed to
preserve recognizable wide, single-pulse GRBs. (For further
information about the sample, one can refer to Norris et al. 1999.)

Only those bursts with the background-subtracted parameters
available are included in our analysis. In addition, we generally
consider the first well identified pulse for each burst since this
pulse is more closely associated with the initial condition of the
event and might be less affected by environment. For each burst we
require that the signal should be detectable at least in three
channels (in this way, the relation between the pulse width and
energy could be studied). With these requirements, we get 62 GRBs
(the KRL sample) from the 67 bursts of the first sample (Kocevski et
al. 2003) and 41 sources (the Norris sample) from the 66 bursts of
the second sample (Norris et al. 1999), respectively. The two
selected samples share the following 19 bursts:
  $\#$563, $\#$914, $\#$1406, $\#$1467, $\#$1883, $\#$2193,
$\#$2387, $\#$2484, $\#$2665, $\#$2880, $\#$3003, $\#$3155,
$\#$3257, $\#$3870, $\#$3875, $\#$3892, $\#$3954, $\#$5517, and
$\#$6504.

The background of light curves is fitted by a polynomial
expression using 1.024 s resolution data that are available from
10 minutes before the trigger to several minutes after the burst.
The data along with the background fit coefficients can be
obtained from the CGRO Science Support Center (CGROSSC) at NASA
Goddard Space Flight Center through its public archives. We adopt
the function presented in equation (22) of Kocevski et al. (2003)
(the KRL function) to fit all of the background-subtracted light
curves since we find that this function could well describe the
observed profile of a FRED pulse. In addition, a fifth parameter
$t_{0}$, which measures the offset between the start of the pulse
and the trigger time, is introduced. The adopted KRL function is
\begin{equation}
F(t)={F_m}(\frac{t+t_0}{t_m+t_0})^r[\frac{d}{d+r}+\frac{r}{d+r}(\frac{t+t_0}{t_m+t_0})^{(r+1)}]^{-\frac{r+d}{r+1}},
\end{equation}
where $t_{m}$ is the time of the pulse's maximum flux, $F_{m}$; r
and d are the power-law rise and decay indexes, respectively. Note
that equation (1) holds for $t\geq -t_0$, when $t< -t_0$ we take
$F(t)=0$.

To obtain an intuitive view on the result of the fit, we develop and
apply an interactive IDL routine for fitting pulses in bursts, which
allows the user to set and adjust the initial pulse parameter
manually before allowing the fitting routine to converge on the
best-fit model via the reduced $\chi^{2}$ minimization. With the two
samples, the fits to the four channel light curves are performed in
sequence for each burst. The fits are examined many times to ensure
that they are indeed the best ones (the reduced $\chi^{2}$ is the
minimum).

We find in our analysis that there are a few with very large
values of the reduced $\chi^{2}$ and a few with very small values.
Shown in Figs. 1 and 2 are typical bursts with very large or very
small values of the reduced $\chi^{2}$ drawn from the two samples,
respectively.
\begin{figure}
\resizebox{7.5cm}{6cm}{\includegraphics{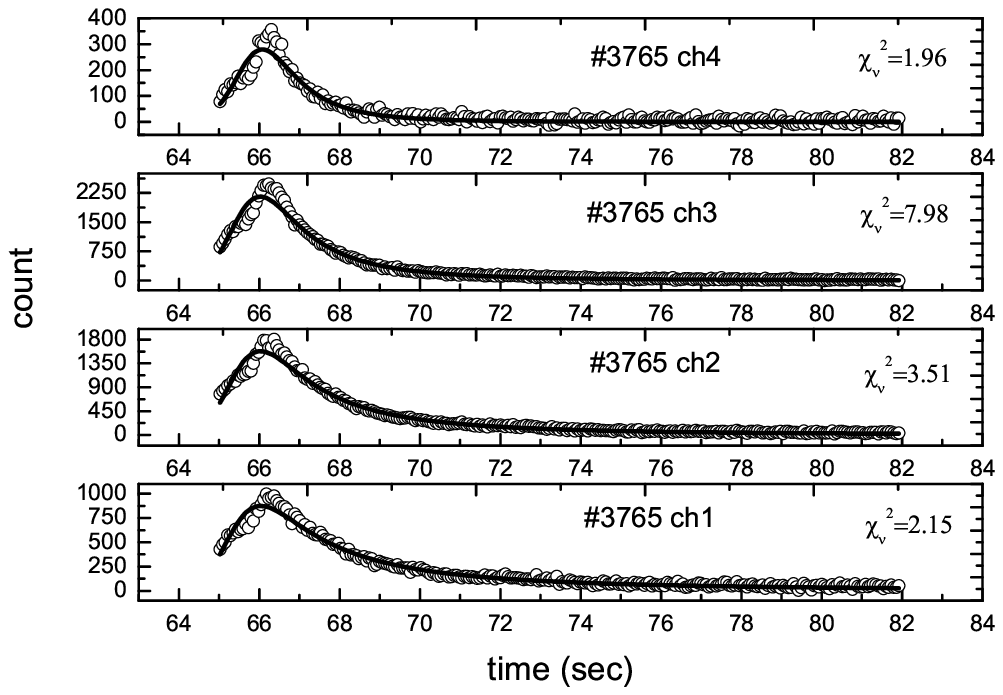}}
\resizebox{7.5cm}{6cm}{\includegraphics{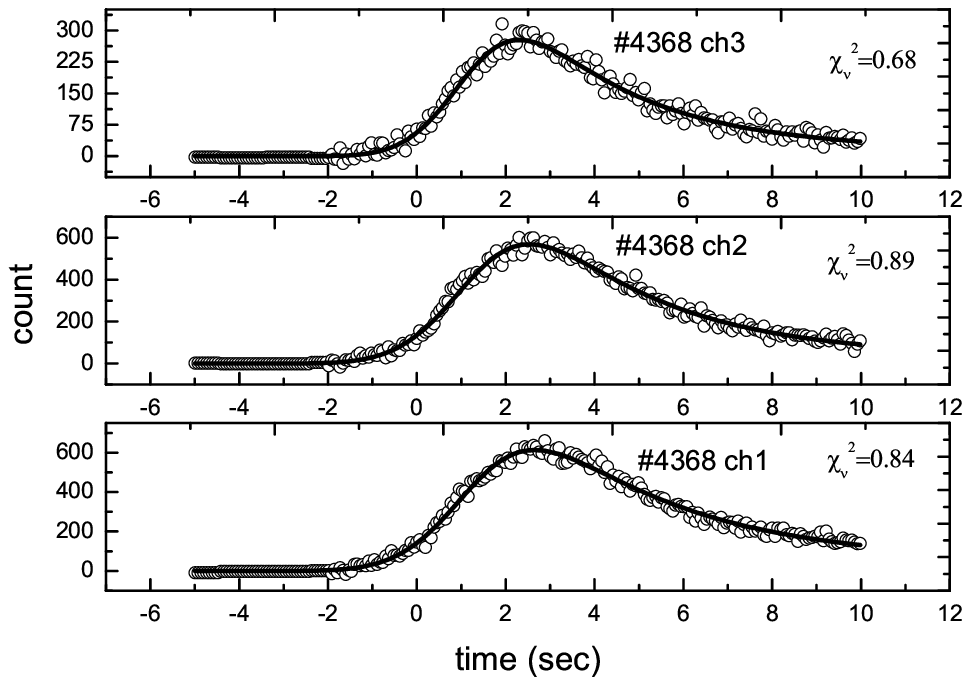}}
\caption{--Plots of the fitting result of a burst with a very large
value of $\chi_{\nu}^{2}$ (left) and a GRB source with a very small
value of $\chi_{\nu}^{2}$ (right) in the KRL sample.}
\end{figure}
\begin{figure}
\resizebox{7.5cm}{6cm}{\includegraphics{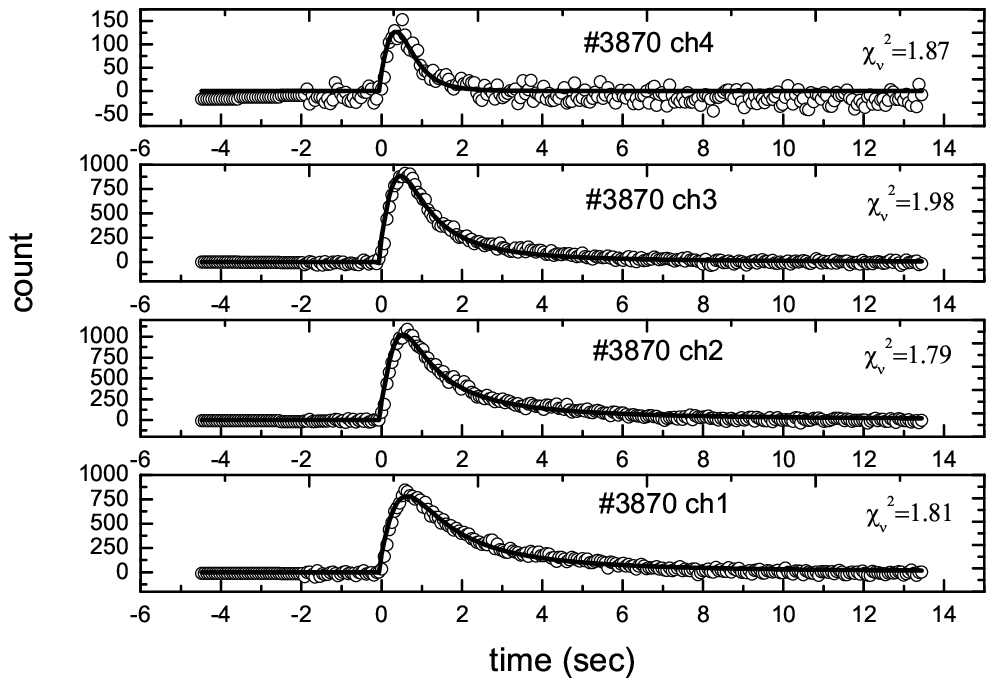}}
\resizebox{7.5cm}{6cm}{\includegraphics{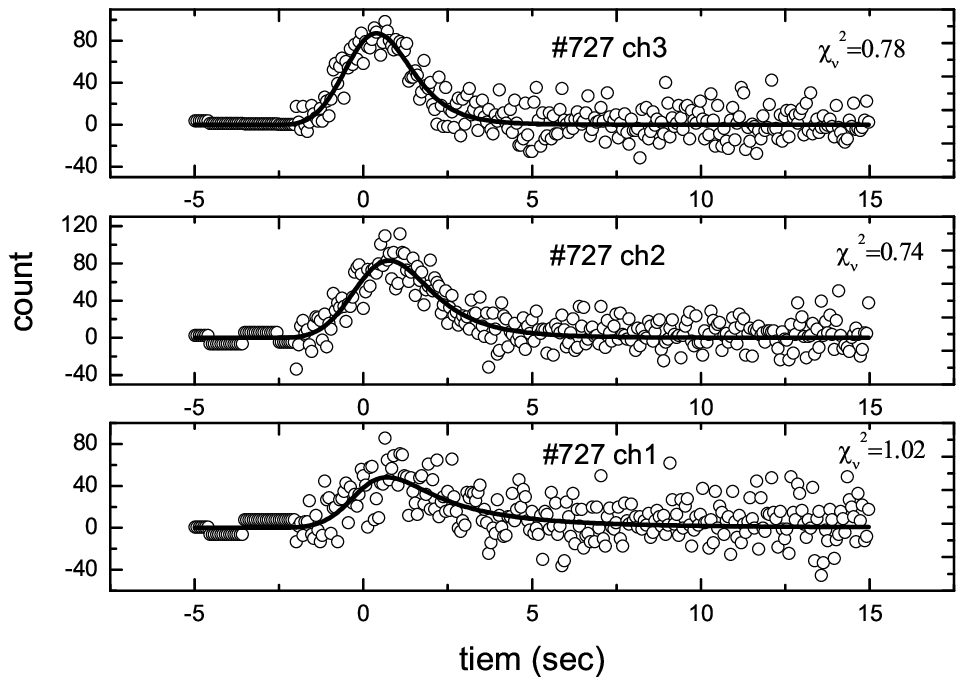}} \caption{--Plots
of the fitting result of a burst with a very large value of
$\chi_{\nu}^{2}$ (left) and a GRB source with a very small value of
$\chi_{\nu}^{2}$ (right) in the Norris sample.}
\end{figure}
One finds from the two figures that, the result of the fit is
quite satisfied. The distributions of the reduced $\chi^{2}$ for
the two samples are displayed in Fig. 3. The reduced $\chi^{2}$
distributions of the KRL and Norris samples peak closely at 1.05
and 1.00, respectively, with a width of approximately 0.200 and
0.200, respectively. The medians for the the KRL and Norris
samples are 1.065 and 1.019, respectively. The distribution of the
reduced $\chi^{2}$, $\chi_{\nu}^{2}$, is so narrow that we are
satisfied by the fits.
\begin{figure}
\centering
\includegraphics[width=5in,angle=0]{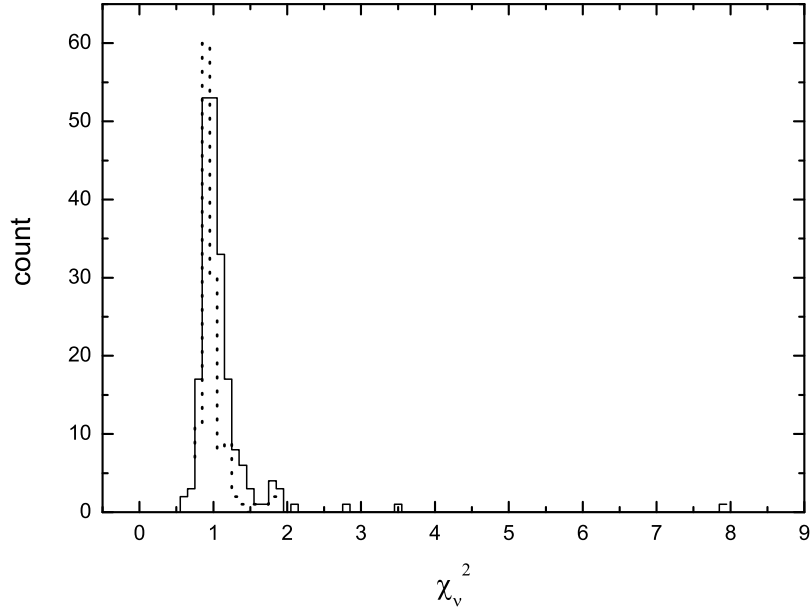}
\caption{--Histograms for the distribution of $\chi_{\nu}^{2}$ in
the KRL sample (the solid line) and the Norris sample (the dash
line). }
\end{figure}

\section{Result}

\subsection{Relationships between the pulse width ratio and energy shown in the two samples}

In the last section we adopt the KRL function described by equation
(1) to fit the light curves of the two samples. The full pulse width
($FWHM$) together with the rising width ($r_{FWHM}$) and the
decaying width ($d_{FWHM}$) in each energy channel for the bursts
are estimated with equation (1) along with the five fitting
parameters of the function. The uncertainties of these widthes are
calculated with the errors of the fitting parameters through the
error transfer formula.

Displayed in Fig. 4 are the distributions of $FWHM$ and
$r_{FWHM}/d_{FWHM}$ (the ratio of the rising width to the decaying
width) for the two samples. We also obtain the relationships
between $FWHM$ and energy and that between $r_{FWHM}/d_{FWHM}$ and
energy for the two samples. Examples of the relationships are
presented in Fig. 5. There are 3 sub-figures (for three bursts)
included in Fig. 5, where each sub-figure is composed of two
panels, with the upper panel showing the plot of $log
FWHM-logE(keV)$ and the lower one displaying the plot of $log
r_{FWHM}/d_{FWHM}-logE(keV)$ for the same source, where $E$ is the
lower energy bound of the four corresponding channels as generally
adopted in previous works (see, e.g., Fenimore et al. 1995).

We find from the distributions of $FWHM$ and $r_{FWHM}/d_{FWHM}$ for
the two samples that: (a) the corresponding values of the medians
are approximately the same for the two samples; (b) while the median
of $FWHM$ declines with energy (when fitting them with a power law,
the index would be negative), $r_{FWHM}/d_{FWHM}$ increases with
energy (when fitting them with a power law, the index would be
positive); (c) the ratio of the rising width to the decaying width
is around 0.5. Combining result (a) and those shown in Figs. 1, 2,
and 3, one can conclude that the KRL and Norris samples share
approximately the same statistical properties. Assuming a power law
relationship, we get from the median of $FWHM$ and energy an index
of -0.267 and obtain from the median of $r_{FWHM}/d_{FWHM}$ and
energy an index of 0.0386 for the KRL sample. The fact that the sign
of the former index is negative and that of the latter is positive
is in well agreement with what predicted previously by Qin et al. in
Paper I. The fact that the median of $r_{FWHM}/d_{FWHM}$ is around
0.5 suggests that the pulses observed are asymmetric, where the
decaying width is generally larger than the rising width by a factor
of two.

From the estimated values of the $FWHM$ and $r_{FWHM}/d_{FWHM}$ of
the four BATSE channels available and the indices derived from
$FWHM$ and energy and from $r_{FWHM}/d_{FWHM}$ and energy for our
selected samples one finds that, for the KRL sample, the largest
values of $r_{FWHM}/d_{FWHM}$ in the four channels are 0.813,
0.699, 0.724, and 0.712 respectively, while the corresponding
values for the Norris sample are 0.723, 0.789, 0.860, and 0.718
respectively. This is in agrement with what predicted previously
in Qin et al. (2004), where they suggested that there is an upper
limit of $r_{FWHM}/d_{FWHM}$ which is approximately 1.3.

Examining the relationship between $FWHM$ and energy and that
between $r_{FWHM}/d_{FWHM}$ and energy for the two samples we find
that there are three kinds of burst identified by the relationship
between $FWHM$ and energy and that between $r_{FWHM}/d_{FWHM}$ and
energy for the two samples. They can be divided by the index of the
power law relationship between $FWHM$ and energy, $\alpha_{FWHM}$,
and the index of the power law relationship between
$r_{FWHM}/d_{FWHM}$ and energy, $\alpha_{ratio}$.

The first class consists of bursts that are seen or suspected to
possess A power law anti-correlation between $FWHM$ and energy
($\alpha_{FWHM}< 0$) and a power law correlation between
$r_{FWHM}/d_{FWHM}$ and energy ($\alpha_{ratio}> 0$). The number is
55 which is around 65\% of the total bursts of the two samples.

The second class includes such bursts that are seen or suspected to
possess negative index power law relationships between $FWHM$ and
energy ($\alpha_{FWHM}< 0$) and between $r_{FWHM}/d_{FWHM}$ and
energy ($\alpha_{ratio}\leq 0$). The number is 24 which is around
29\% of the total bursts of the two samples.

The third class contains the bursts that are seen or suspected to
possess a power law correlation between $FWHM$ and energy
($\alpha_{FWHM}\geq 0$). The number is 5 which is around 6\% of the
total bursts of the two samples.

Bursts of class 1 are obviously those predicted in Paper I, in
which the prediction was performed under the assumption that the
so-called curvature effect (the Doppler and time delay effect over
the relativistically expanding fireball surface) is important. The
fact that the number of bursts of this class is around 65\% of the
total bursts of the two samples indicates that light curves of the
majority of bursts of these samples are likely to suffer from the
curvature effect. This conclusion is in agreement with what
suggested in Qin \& Lu (2005) and Paper I. Sources of the second
class are not predicted in Paper I. Recalled that what discussed
in Paper I involves only simple cases. For example, when they
considered the situation of the rest frame radiation varying with
time, they discussed only a varying Band function and involved
only a pattern of a linear softening. It is unclear if a varying
synchrotron or Comptonized radiation is concerned or a different
pattern of the spectral evolution is considered, both negative
index power law relationships for a burst could be expected (this
might deserves a further investigation). For the same reason, the
mechanism accounting for the third class is also unclear.

\subsection{Relationship between power law indices}

As predicted in Paper I and suggested in the bursts of class 1, the
power law phase in the relationship between $r_{FWHM}/d_{FWHM}$ and
energy shows an opposite tendency to that between $FWHM$ and energy.
We are curious about how the two power law indices $\alpha_{FWHM}$
and $\alpha_{ratio}$ being related. Presented in Fig. 6 is the plot
of $\alpha_{ratio}$ vs. $\alpha_{FWHM}$ for the KRL and Norris
samples. An anti-correlation between the two quantities is observed
in both samples. For the KRL sample, a regression analysis yields
$\alpha_{ratio}=(-0.03\pm 0.03)-(0.23\pm 0.06)\alpha_{FWHM}$, with
the correlation coefficient being $r = -0.428$ ($N = 62$), while for
the Norris sample, the analysis produces $\alpha_{ratio}=(-0.01\pm
0.04)-(0.29\pm 0.10)\alpha_{FWHM}$, with the correlation coefficient
being $r = -0.425$ ($N = 41$). Plotted in Fig. 6 are also three
distribution regions for the two indices, which are associated with
the three classes defined above. Since for a certain energy range
the sign of the indices depends obviously on the radiation mechanism
(see Figs. 1, 2 and 3 in Paper I), these different regions might
correspond to different mechanisms (this deserves a detailed
analysis).

Distributions of $\alpha_{FWHM}$ and $\alpha_{ratio}$ are
displayed in Fig. 7. For the KRL sample, the median of the
distribution of $\alpha_{FWHM}$ is $-0.277$ and that of
$\alpha_{ratio}$ is $0.066$. For the Norris sample, the medians of
the distributions of $\alpha_{FWHM}$ and $\alpha_{ratio}$ are
$-0.302$ and $0.083$, respectively. Two statistical
characteristics are observed. One is that the typical value of
$\alpha_{FWHM}$ is negative while that of $\alpha_{ratio}$ is
positive. The other is that the absolute value of the typical
$\alpha_{FWHM}$ is about four times of that of $\alpha_{ratio}$.
If this is expectable by the Doppler model is unclear. In
addition, we find that the distribution of $\alpha_{ratio}$ is
much narrower than that of $\alpha_{FWHM}$, which suggests that if
served as a parameter associated with mechanisms, the latter index
must be more sensitive than the former.


\section{Discussion and conclusions}

As predicted previously, emission of pulses over a relativistically
expanding fireball surface could lead to A power law
anti-correlation between the pulse width and energy and a power law
correlation between the ratio of the rising width to the decaying
width and energy. Although A power law anti-correlation between the
pulse width and energy was observed by many authors, it is unclear
if a power law correlation between the width ratio and energy could
be detected in the same sources. In this paper we investigate this
issue with two samples which contain well identified pulses, with
one being the KRL sample (Kocevski et al. 2003) and the other being
the Norris sample (Norris et al. 1999).

There are 84 sources in total for the two samples (where 19 bursts
are included in both samples). Shown in these samples, a power law
relationship could indeed be well established between not only the
pulse width and energy but also the ratio of rising width to the
decaying width and energy. A power law anti-correlation between
$FWHM$ and energy and a power law correlation between
$r_{FWHM}/d_{FWHM}$ and energy are seen in the light curves of the
majority (around 65\%) of bursts of the two samples. This suggests
that these bursts are likely to arise from the emission associated
with the shocks occurred on a relativistically expanding fireball
surface, where the curvature effect must be important (see Paper I).
For the rest of the bursts, the corresponding mechanism is currently
unclear. We propose that a varying synchrotron or Comptonized
radiation or a different pattern of the spectral evolution should be
concerned. In this case, one might be sure if the observed
relationships for these rest bursts can also be accounted for by the
curvature effect.

In addition, we find that the largest values of $r_{FWHM}/d_{FWHM}$
in the four channels of the two samples do not exceed 0.9, which is
in agrement with what predicted previously in Qin et al. (2004),
where they suggested that there is an upper limit of
$r_{FWHM}/d_{FWHM}$ which is approximately 1.3.

An analysis of the relationship between the two power law indices
$\alpha_{FWHM}$ and $\alpha_{ratio}$ reveals an anti-correlation
between the two. We divide the $\alpha_{ratio} - \alpha_{FWHM}$
plane into three regions. They are regions I ($\alpha_{FWHM}< 0$
and $\alpha_{ratio}> 0$), II ($\alpha_{FWHM}< 0$ and
$\alpha_{ratio}\leq 0$) and III ($\alpha_{FWHM}\geq 0$) (see Fig.
6). Sources inside these regions are defined as classes 1, 2 and
3, respectively. While bursts in region I (class 1) were predicted
previously, those in regions II (class 2) and III (class 3) are
unfamiliar. We suspect that different locations of
($\alpha_{FWHM}$, $\alpha_{ratio}$) might correspond to different
mechanisms such as the pattern of the evolution and the real form
of the rest frame spectrum. If so, the plot of $\alpha_{ratio} -
\alpha_{FWHM}$ might be useful to provide information of
mechanisms. We observe that the absolute value of the typical
$\alpha_{FWHM}$ is about four times of that of $\alpha_{ratio}$,
and the distribution of $\alpha_{ratio}$ is much narrower than
that of $\alpha_{FWHM}$. This indicates that, if they are
parameters confined by mechanisms, $\alpha_{FWHM}$ must be more
sensitive than $\alpha_{ratio}$.

One might observe from the estimated values of the $FWHM$ and
$r_{FWHM}/d_{FWHM}$ of the four BATSE channels available and the
indices derived from $FWHM$ and energy and from
$r_{FWHM}/d_{FWHM}$ and energy for the KRL sample and the Norris
sample that uncertainties of $\alpha_{ratio}$ are larger than the
uncertainties of the corresponding index $\alpha_{FWHM}$. This is
due to the error transform nature (note that the uncertainty of
$\alpha_{FWHM}$ is determined by the uncertainty of
$r_{FWHM}+d_{FWHM}$, while the uncertainty of $\alpha_{ratio}$ is
determined by the uncertainty of $r_{FWHM}/d_{FWHM}$; the later
must be larger than the former). The large uncertainty and the
narrow distribution makes the estimated values of $\alpha_{ratio}$
quite uncertain. This might misidentify some bursts of class 1 as
those of class 2, or vice versa. Therefore, definitions of many of
the bursts of classes 1 and 2 are not certain, and thus the
percentage of the number of any of the classes to the total number
is not certain. However, due to the following reasons we argue
that this is unlikely to change the percentage dramatically. The
first is that we have checked each burst very carefully and then
have been sure that the fitting curves pass through indeed the
central regions of the observed data. This could be confirmed by
the very narrow distribution of the reduced $\chi^{2}$ shown in
Fig. 3 (for the goodness of fit one can also refer to Figs. 1 and
2). The second is that while some bursts of class 1 might be
misidentified as those of class 2, some sources of class 2 might
also be misclassified as those of class 1, and this will ease the
problem (one can observe from Fig. 6 that there are bursts of both
classes 1 and 2 located around the horizon line of
$\alpha_{ratio}=0$).

As suggested by Qin et al. in Paper I, for the two relationships
concerned, there would be a plateau or a slope (appeared also as a
power law) beyond the main power law range, depending on the form
and the evolution pattern of the rest-frame spectrum. These were
noticed in the much smaller sample employed in Paper I. They are
also observed in the two samples employed here. It should be noticed
that in some cases these features might lead to a smaller absolute
value of the power law index (e.g., when the turnover appears within
the energy range concerned). If the energy range of observation is
large enough, one can expect to measure the indices within the main
power law range for each burst, and in that case the bursts would be
easier to classified. Also in this case the lower and upper limits
of the main power law range would be well measured and this in turn
would provide an independent test to the Doppler model. As revealed
in Paper I, besides the common features (the plateau/power
law/plateau and the peaked features), there exhibit other features
in a few cases. For a small number of bursts in our samples, a
abnormal sinkage feature could be observed in the two relationships,
which is not a result predicted in Paper I. What causes this is
unclear.

The conclusion that the ratio of the rising width to the decaying
width of the majority of bursts tend to be larger at higher energies
shown in this paper is conflicted with what was noticed previously.
We argue that this effect is indeed very small and is hard to be
observed as pointed out above. However, this tendency holds in terms
of statistics. The tendency can also be observed from Fig. 7 (right
panel), where the majority of bursts have $\alpha_{ratio}$ greater
than zero. Direct evidence of the tendency can be obtained in the
relationships between $FWHM$ and energy and that between
$r_{FWHM}/d_{FWHM}$ and energy for the two samples, when one paying
attention to the bursts of class 1, which are about 65\% of the
total bursts. We suspect that it is the small absolute values of
$\alpha_{ratio}$ that make the detection difficult and this probably
leads to the un-detection of the tendency in previous works.

As suggested previously, when the opening angle of uniform jets is
sufficiently larger (say, much larger than $1/\Gamma$), the pulse
observed would not be significantly different from that arising
from the whole fireball surface (see Qin and Lu 2005). Therefore,
conclusions favoring a fireball generally favor a uniform jet.


This work was supported by the Special Funds for Major State Basic
Research Projects (``973'') and National Natural Science
Foundation of China (No. 10273019).

\newpage

\begin{figure*}
\resizebox{7.5cm}{!}{\includegraphics{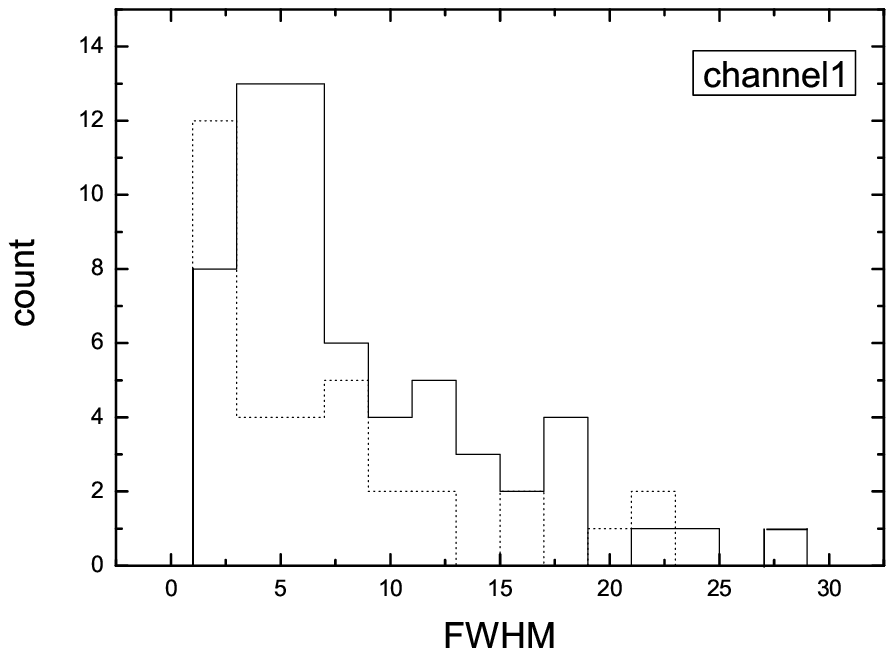}}
\resizebox{7.5cm}{!}{\includegraphics{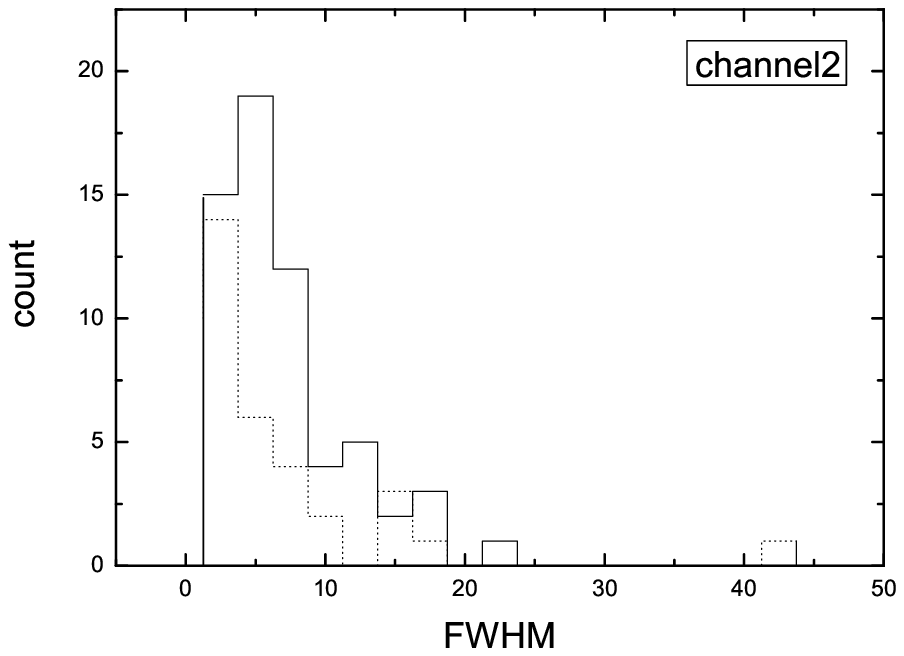}}
\resizebox{7.5cm}{!}{\includegraphics{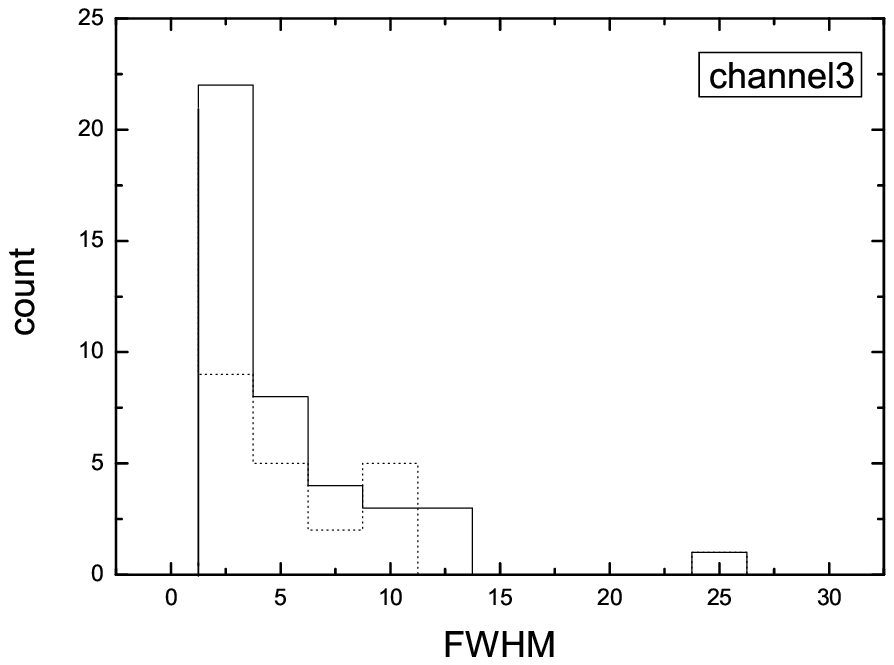}}
\resizebox{7.5cm}{!}{\includegraphics{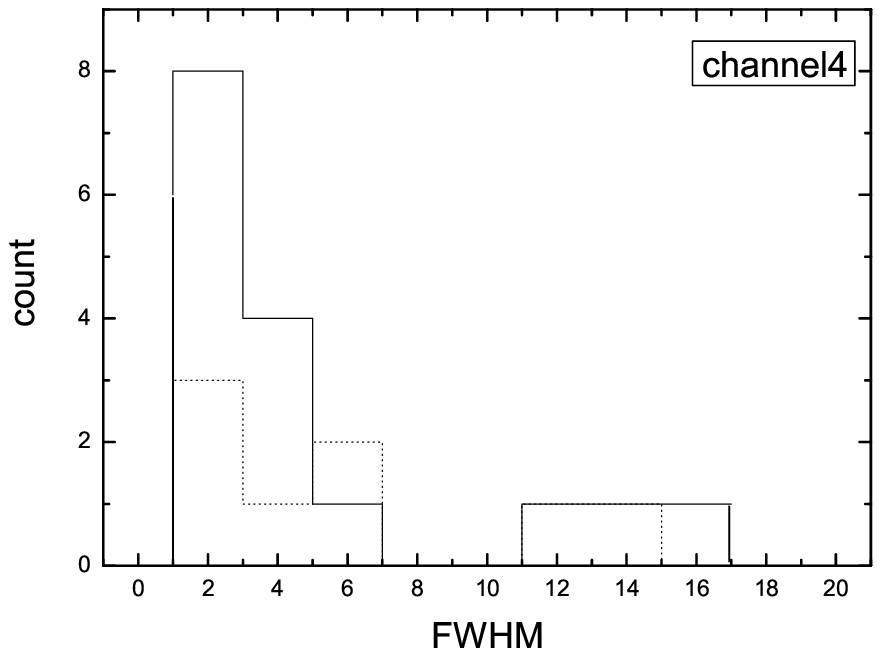}}
\resizebox{7.5cm}{!}{\includegraphics{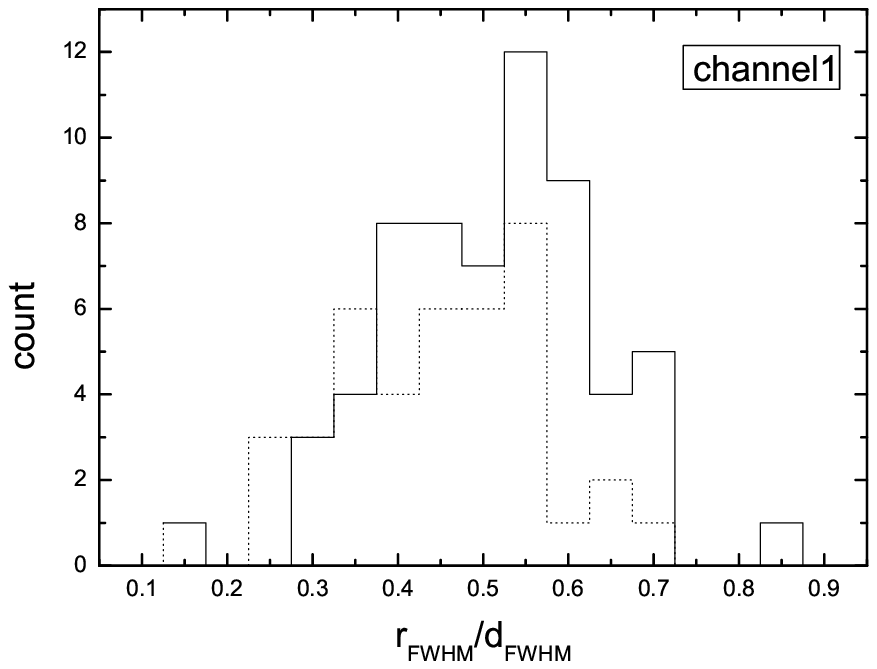}}
\resizebox{7.5cm}{!}{\includegraphics{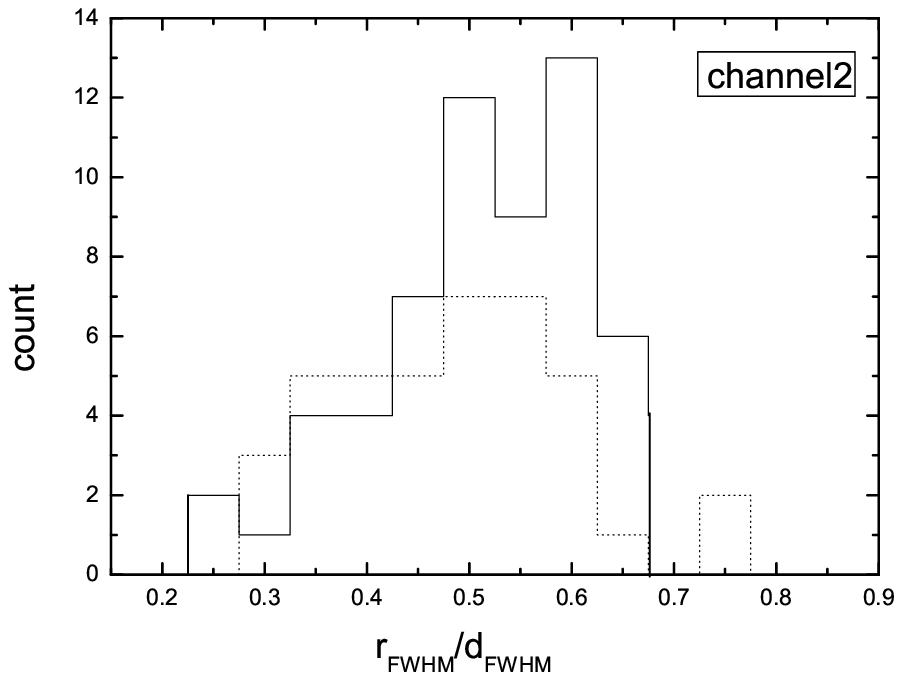}}
\resizebox{7.5cm}{!}{\includegraphics{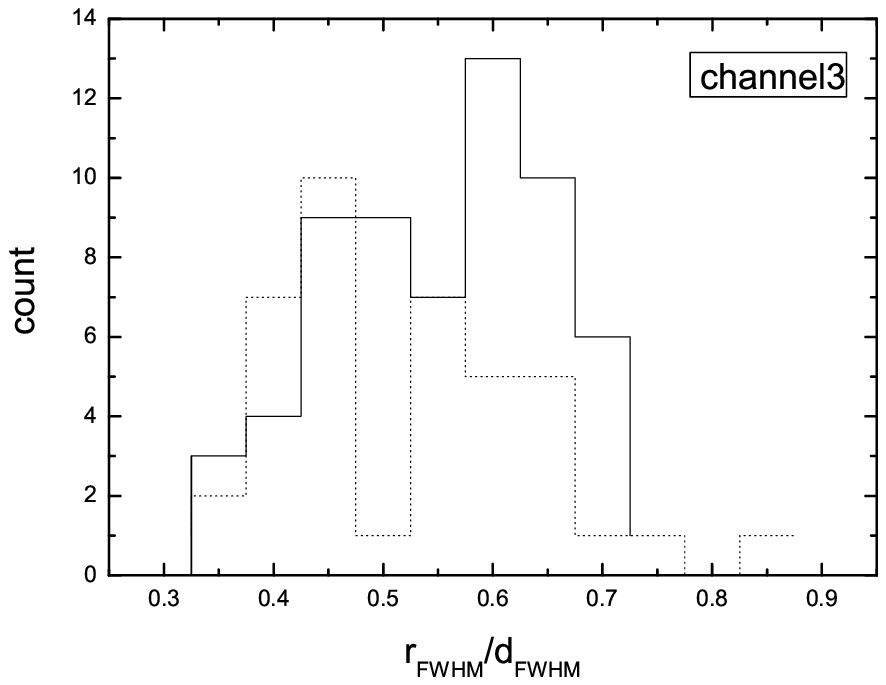}}
\resizebox{7.5cm}{!}{\includegraphics{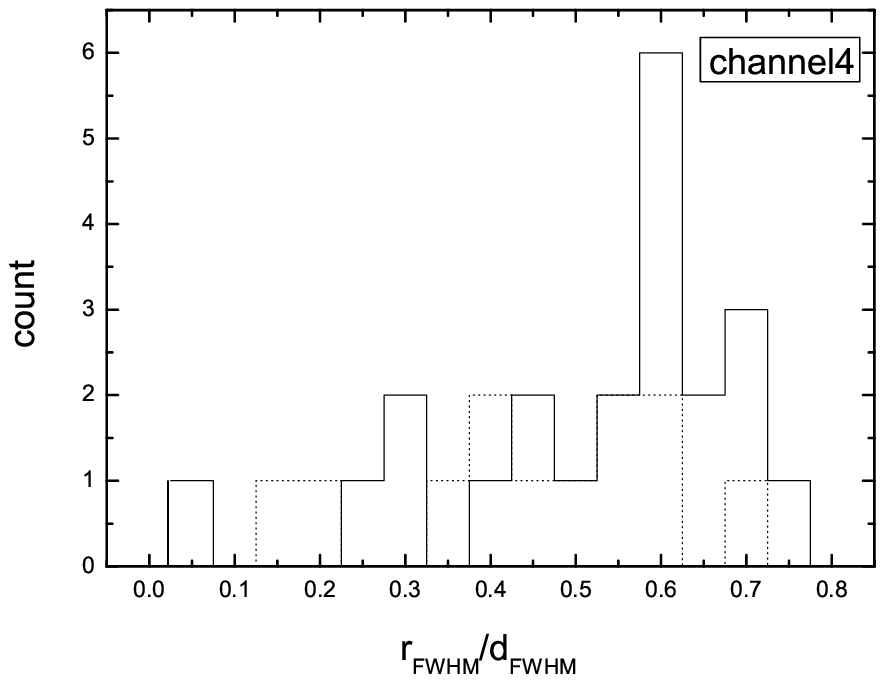}}
 \caption{Distributions of $FWHM$ (the 4 upper panels) and
$r_{FWHM}/d_{FWHM}$ (the 4 lower panels) in the four energy channels
for the KRL sample (solid lines) and the Norris sample (dot lines),
respectively. }
\end{figure*}

\begin{figure*}
 \resizebox{5cm}{5cm}{\includegraphics{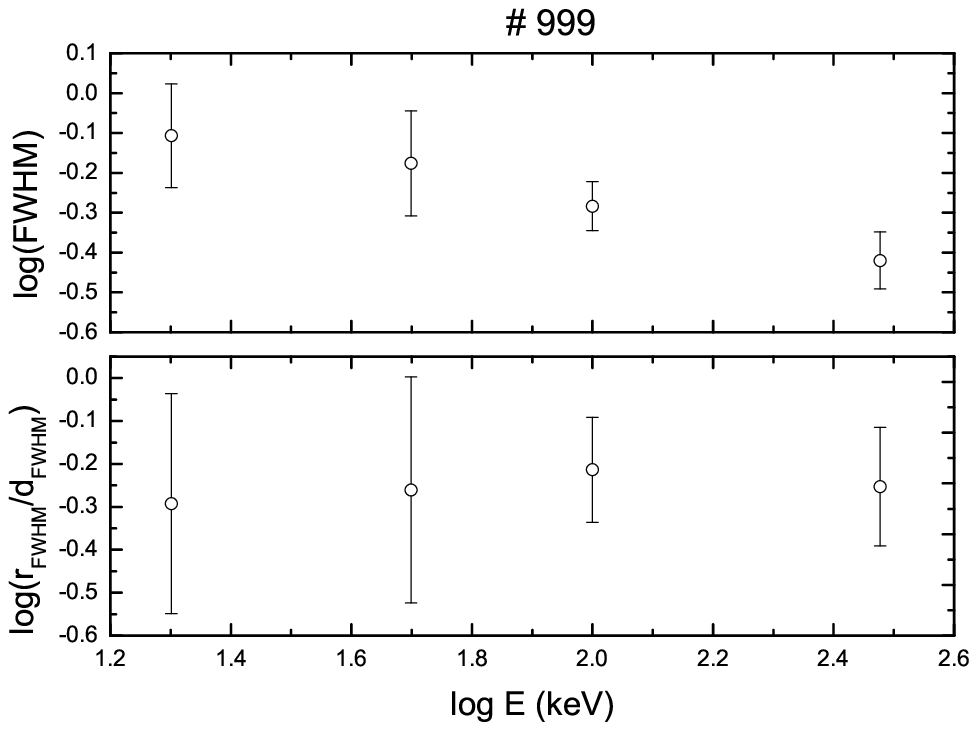}}
 \resizebox{5cm}{5cm}{\includegraphics{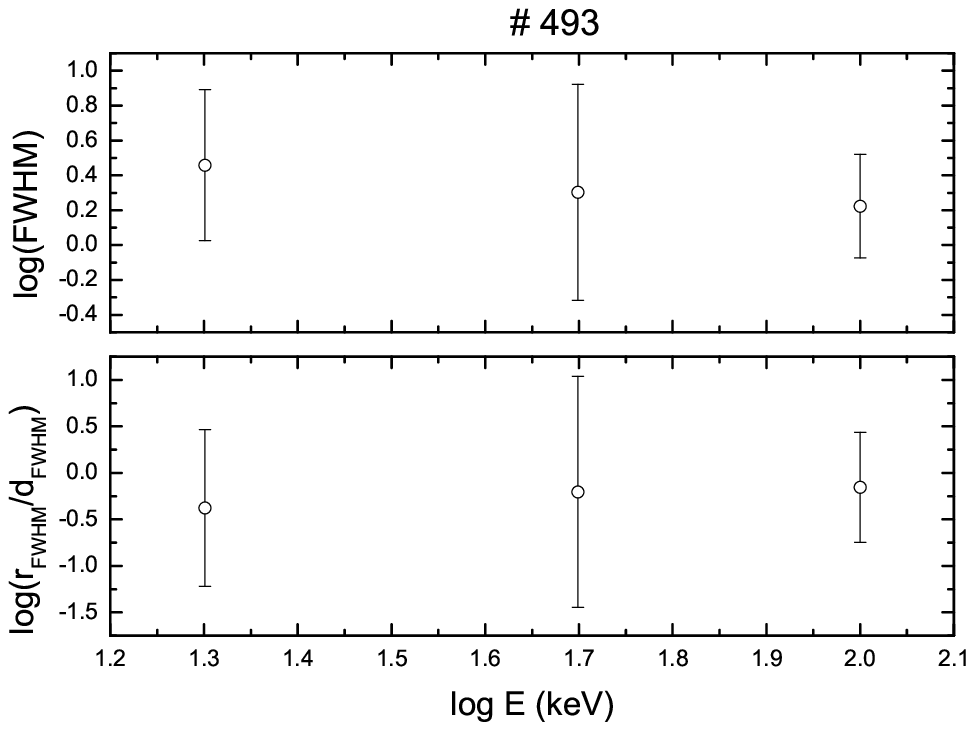}}
 \resizebox{5cm}{5cm}{\includegraphics{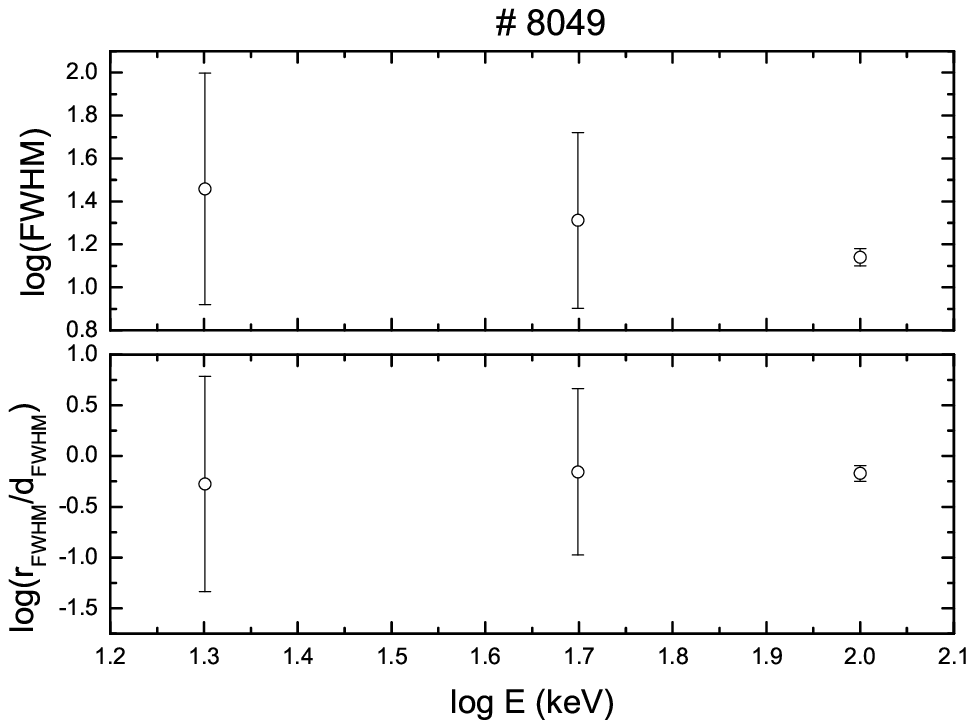}}
\caption{--Example plots of the relationship between the observed
$FWHM$ width of pulse and energy (the upper panel) and that
between the ratio of the $FWHM$ width of the rising portion to
that of the decaying phase of the light curve of pulses and energy
(the lower panel) for our selected samples.} \label{sed}
\end{figure*}

\begin{figure}
\centering
\resizebox{7.5cm}{!}{\includegraphics{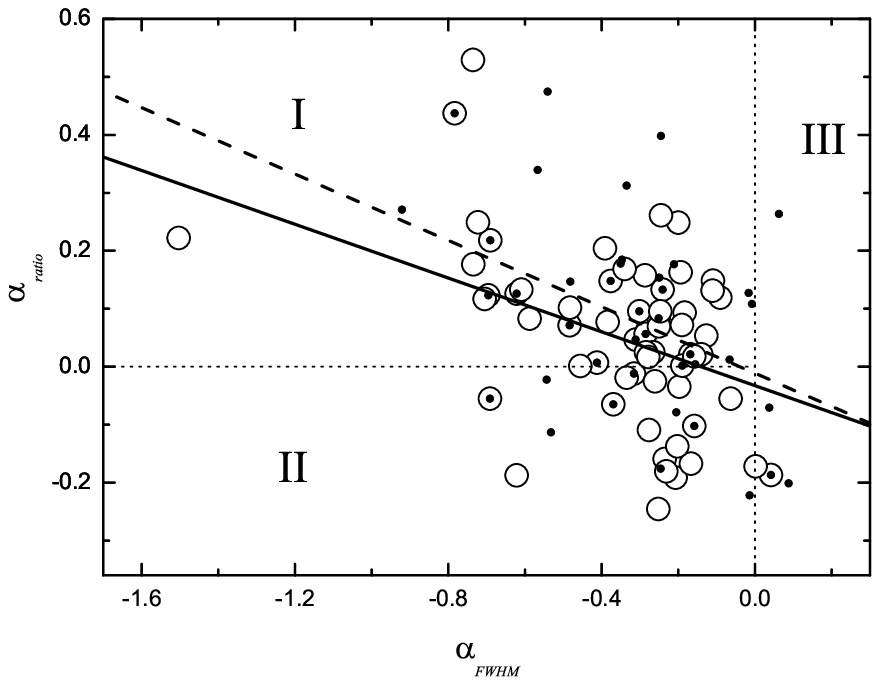}}
\resizebox{7.5cm}{!}{\includegraphics{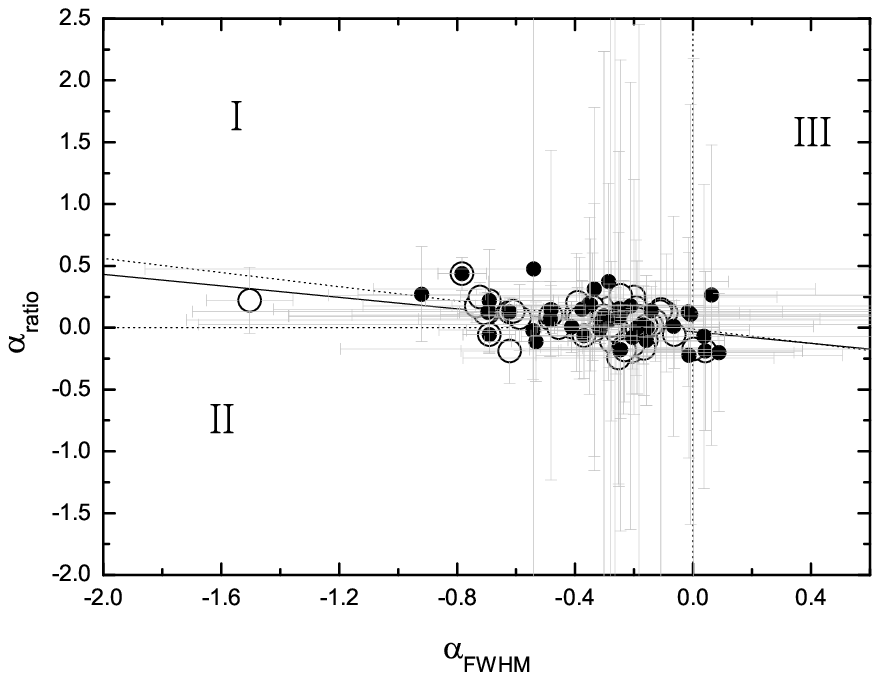}}
\caption{--Relation between the two indices obtained from $FWHM$ and
energy and the ratio of the $FWHM$ width of the rising portion to
that of the decaying phase of the light curve of pulse and energy
with no error bars (the left panel) and that with error bars (the
right panel), where the open circles present the KRL sample and the
filled circles stand for the Norris sample. The solid line and the
dot line are the regression lines for the KRL sample and the Norris
sample, respectively.}
\end{figure}

\begin{figure}
\centering \resizebox{7.5cm}{6cm}{\includegraphics{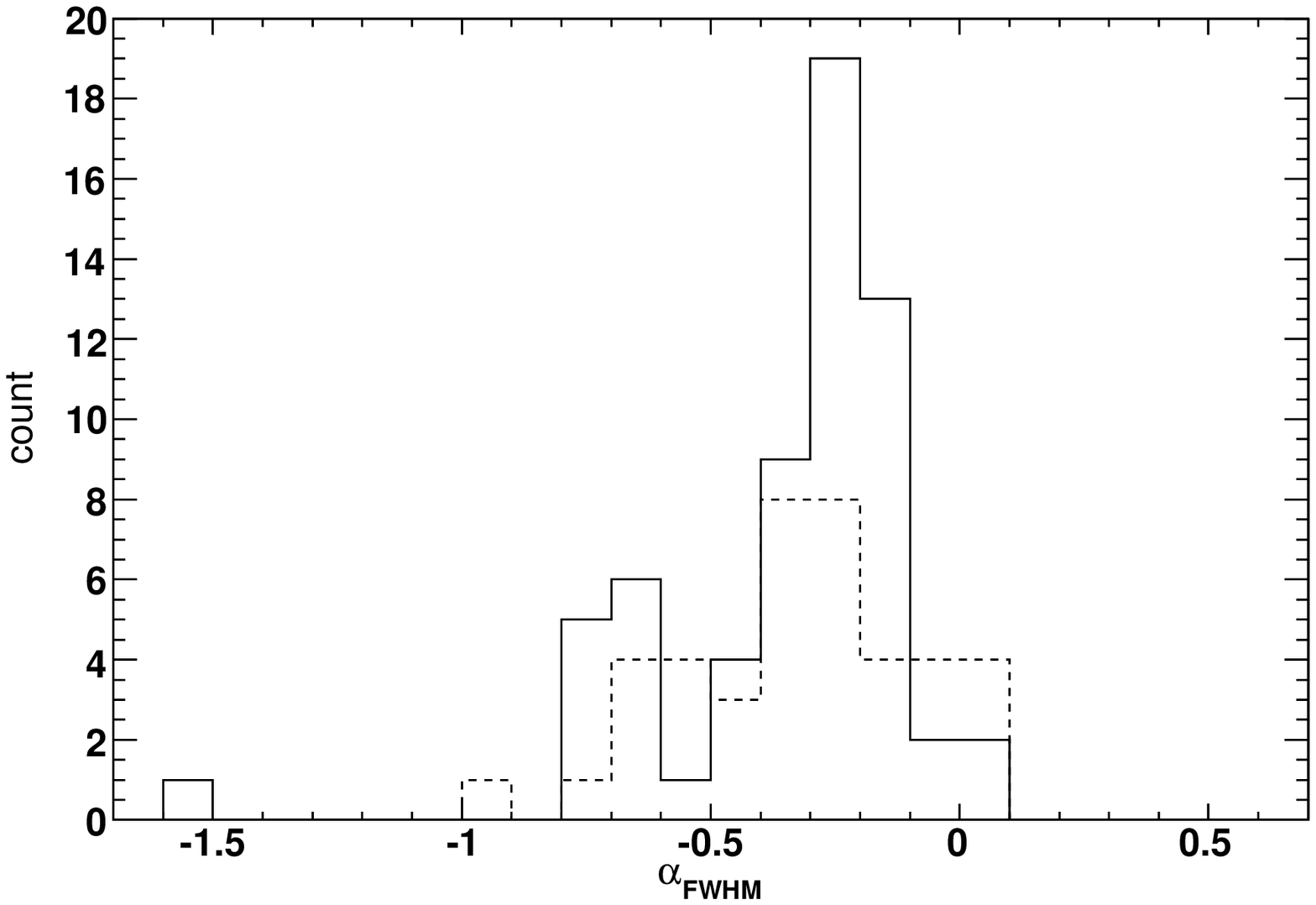}}
\resizebox{7.5cm}{6cm}{\includegraphics{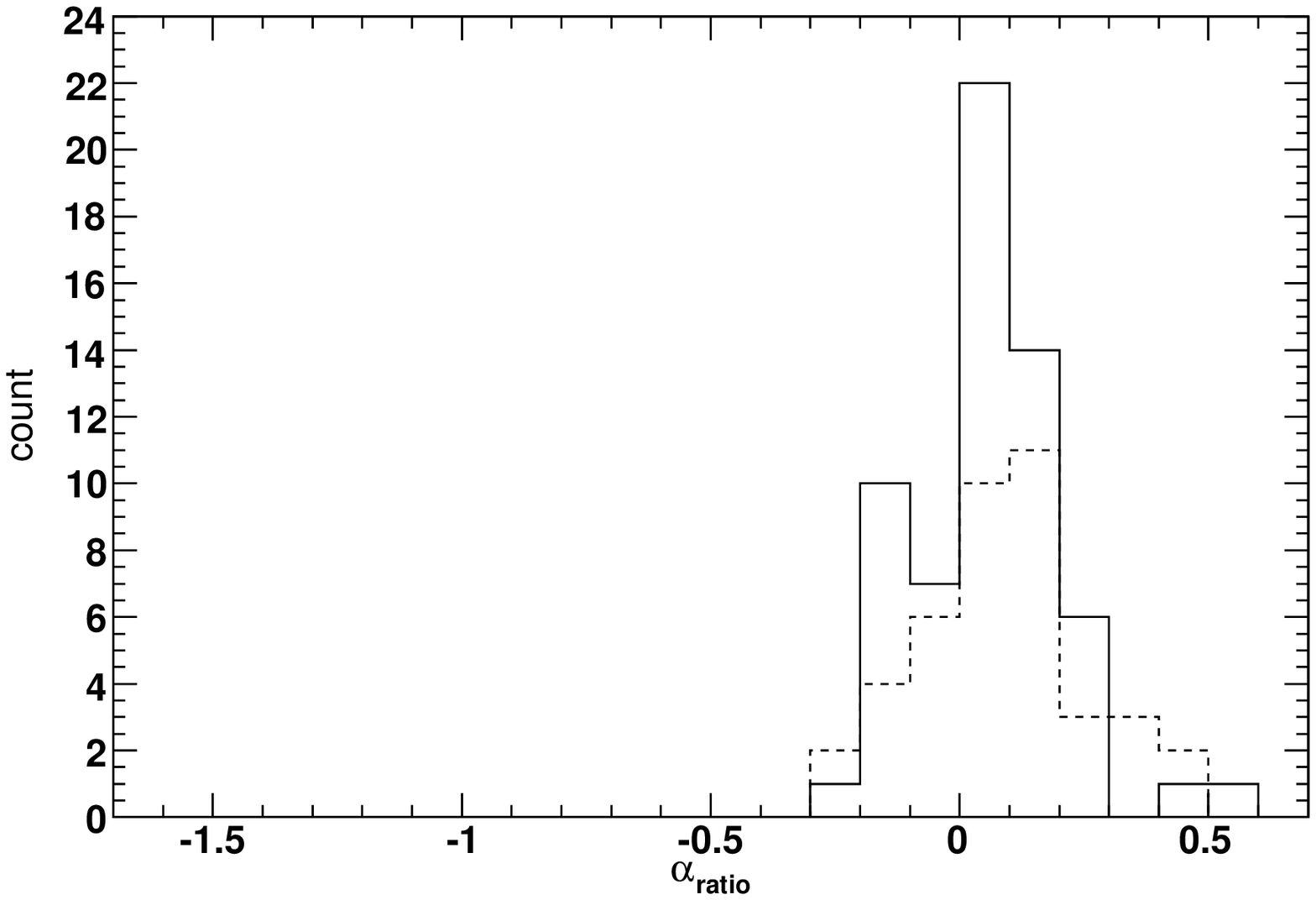}}
\caption{--Histograms for the distributions of the indices obtained
from FWHM and energy (the left panel) and the ratio of the FWHM
width of the rising portion to that of the decaying phase of the
light curve of pulse and energy (the right panel), where the solid
line presents the KRL sample and the dash line stands for the Norris
sample, respectively. }
\end{figure}

\newpage

\label{lastpage}

\end{document}